\documentclass{article}
\usepackage{amsmath}
\usepackage{amssymb}
 \newtheorem{defi}{Definition}
 \newtheorem{lem}{Lemma}
 \newtheorem{thm}{Theorem}

\makeatletter
\newenvironment{pf}{\noindent{\bf Proof}\quad}
{\leavevmode\hfill$\square$\par\@endpetrue}
\makeatother

\def\tr{\mathop{\rm tr}\nolimits}
\def\SU{\mathop{\rm SU}\nolimits}
\def\id{\mathop{\rm Id}\nolimits}
\def\Pr{\mathop{\rm Pr}\nolimits}
\def\Otimes{\mathop{\otimes}}
\def\pro{{\cal P}({\cal H})}
\def\proa{{\cal P}({\cal H}')}
\begin{document}
\vskip 2em \begin{center}
 {\LARGE\bf 
 Asymptotic estimation theory \par
for a finite dimensional pure state model
\par} \vskip 1.5em 
\large \lineskip .5em
Masahito Hayashi \par
Department of Mathematics, Kyoto University \par
 Kyoto 606-8502, Japan \par 
e-mail address: masahito@kusm.kyoto-u.ac.jp \par
\end{center}
\begin{abstract}
The optimization of measurement for $n$ samples of pure sates are studied.
The error of the optimal measurement for $n$ samples
is asymptotically compared with the one of the maximum 
likelihood estimators from $n$
data given by the optimal measurement for one sample. 
\end{abstract}
\section{Introduction}
\indent Recently, there has been a rise in the necessity for studies about 
statistical estimation for the unknown state,
related to the corresponding 
advance in measuring technologies in quantum optics.
An investigation including both quantum theory 
and mathematical statistics is necessary for an essential 
understanding of quantum theory
because it has statistical aspects \cite{Hel,HolP}.
Therefore, it is indeed important to optimize the measuring process 
with respect to the estimation of the unknown state.
Such research is known as quantum estimation, and
was initiated by Helstrom in the late 1960s,
originating in
the optimization of the detecting process in 
optical communications \cite{Hel}.
In classical statistical estimation, 
one searches the most suitable estimator
for which probability measure 
describes the objective probabilistic phenomenon.
In quantum estimation, one searches the most suitable measurement for
which density operator describes the objective quantum state.

Contained among important results are three estimation problems.
The first is of the complex amplitude of coherent light in thermal noise 
and the second is of the expectation parameters of quantum Gaussian state.
The former was studied by Yuen and Lax \cite{YL}
and the latter by Holevo \cite{HolP}.
These studies discovered that heterodyning is the most suitable for
the estimation of the complex amplitude of coherent light in thermal noise.
The third is a formulation of the covariant measurement with respect to
an action of a group. It was studied by Holevo \cite{HolP,Holc}.
In the formulation, he established a quantum analogue of Hunt-Stein theorem.

Quantum estimation, was first used in the evaluation of the 
estimation error of a single sample of the unknown state as it 
had advanced in connection with the optimization of the measuring process 
in optical communications.
Thus early studies were lacking in asymptotic aspects,
i.e. there were few researches with respect to
reducing the estimation error by quantum correlations between samples. 

Recently, studies about the estimation of the unknown state
are attracting many physicists \cite{Jon,Jon1,Da,DY}.
Some of them were drawn by the variation of the measuring precision
with respect to the number of samples of the unknown state \cite{BAD,smsp}.

Nagaoka \cite{Na1} studied, for the first time,
asymptotic aspects of quantum estimation.
He paid particular attention to the quantum correlations 
between samples of the unknown state,
and studied the relation between
the asymptotic estimation and the local detection of a one-parameter family
of quantum states. 

In the early 1990s, Fujiwara and Nagaoka \cite{F1,FN2,FN} studied 
the estimation problem for a multi-parameter family consisting of pure states.
They pioneered studies into the estimation problem of the complex amplitude
of noiseless coherent light.
The research found that heterodyning is the most suitable
for the estimation of the complex amplitude of noiseless
coherent light as for the one of coherent light in thermal noise.
In 1996, Matsumoto \cite{Matsu} established a more general formulation of
the estimation for a multi-parameter family consisting of pure states.
Moreover in 1991, Nagaoka \cite{Naga1} treated the estimation problem
for 2-parameter families of mixed states in spin 1/2 system,
and in 1997 the author \cite{Haya1,Haya2} treated it for 3-parameter 
families of mixed states in spin 1/2 system.
However, there are no asymptotic aspects in these works 
about multi-parameter families.
There is more necessity of this type of investigation 
into one- and multi-parameter families.

Can quantum estimation reduce the estimation error
by using the quantum correlations between samples, 
under the preparation of sufficient samples of the unknown state?
To answer this question, in this paper, we treat a family,
consisting of all of pure states on a
Hilbert space ${\cal H}$
\footnote{Where ${\cal H}$ denotes a finite-dimensional 
Hilbert space which corresponds to the physical system of interest.}
under the preparation of $n$ samples of the unknown state,
with the estimation problem 
In \S \ref{pure1}, we use, as a tool, the 
composite system consisting of $n$ samples as a single system.
The quantum i.i.d. condition is introduced as the quantum counterpart of the
independent and identical distributions condition (\ref{iid4}).
In \S \ref{gun1}, we review Holevo's result concerning
covariant measurements which will be used in the following sections.
In \S \ref{saiteki}, we apply Holevo's result to the optimization
of measurements on the composite system,
which results in obtaining the most suitable measurement (Theorem \ref{main}).
We asymptotically calculate the estimation error by the optimal
measurement in the sense of both the mean square error (MSE) and large
deviation.
(see (\ref{keha})(\ref{bun1})(\ref{zenkin3})
(\ref{zenkin5}).)
The first term of the right-hand side of (\ref{bun1}) is consistent with the 
value conjectured from the results
by Fujiwara, Nagaoka \cite{FN} and Matsumoto \cite{Matsu}.
However, the optimal measurement may be too difficult for modern technology
to realize when using more than one samples.

In \S \ref{fish}, we use this estimation problem
under the following guidelines.
The samples are divided into pairs consisting
of a maximum of $m$ samples.
By measuring each pair with the optimal measurement in section \ref{saiteki},
we create some data.
The estimated valued is given by manipulating these data. 
The restricted condition is called $m$-semiclassical
(see (\ref{semic})).
We compare an $m$-semiclassical measurement with the optimal 
measurement of section \ref{saiteki}
with respect to the estimation error under the preparation  
of a sufficient amount of samples.
When we use the maximum likelihood estimator to manipulate the data,
the MSE of both asymptotically coincide in the first order
(see (\ref{bun1})(\ref{bun2})).
However, when the radius of allowable errors is finite,
the error of large deviation type in the latter is smaller than 
that in the former type
(see (\ref{zenkin3})(\ref{div2})).

Can we asymptotically realize a small estimation error as 
the optimal measurements in section \ref{saiteki} has?
It is, physically, sufficient
to construct the optimal measurement for one sample.
In section \ref{fish}, we show how to construct it
(see (\ref{hor})).
And in this case, we can calculate the maximum likelihood estimator
from data by using computer.

Most of the proofs of this paper are given in Appendices.
In view of multiparameter families of mixed states in spin 1/2 system,
Hayashi \cite{Haya3} has discussed the same problem 
by using Cram\'{e}r-Rao type bound.

\section{Pure state $n$-i.i.d. model}\label{pure1}
In this section, we use the mathematical formulation 
of the estimation for pure states.
Let $k$ be the dimension of the Hilbert space ${\cal H}$, and
$\pro$ be the set of pure states on ${\cal H}$.

In quantum physics, the most general description of a quantum measurement 
is given by the mathematical concept of 
a {\it positive operator valued measure} (POVM) \cite{Hel,HolP} 
on the system of state space.
Generally, if $\Omega$ is measurable space, 
a measurement $M$ satisfies the following:
\begin{eqnarray*}
M(B)=M(B)^*, M(B) \ge 0, M( \emptyset ) = 0, M(\Omega) = 
\id \hbox{ \rm on } {\cal H},
\hbox{ \rm for any } B \subset \Omega. \\
M( \cup_i B_i ) = \sum_i M( B_i), \hbox{ \rm for } B_i \cap B_j = \emptyset
(i \neq j), \{ B_i \} \hbox{ \rm is countable subsets of } \Omega.
\end{eqnarray*}
In this paper, ${\cal M}(\Omega,{\cal H})$ denotes the set of 
POVMs on ${\cal H}$ whose measurable set is $\Omega$.
A measurement $M \in {\cal M}(\Omega,{\cal H})$ is called simple
if $M(B)$ is a projection for any Borel $B \subset \Omega$.
A measurement $M$ is random if
it is described as a convex combination of simple measurements.
A random measurement $M = \sum_i a_i M_i$ ($M_i$ is simple and $a_i \,> 0$.) 
can be realized 
when every measurement $M_i$ is done with the probability $a_i$. 

In this paper, we consider measurements whose measurable set is 
$\pro$
since it is known that the unknown state is included in ${\cal P}({\cal H})$.

Next, we define two distances charactering the homogeneous space $\pro$.
\begin{defi}\label{fsb}
 the Fubini-Study distance $d_{fs}$ (which is the geodesic distance of the Fubini-Study metric) is defined as:
\begin{eqnarray}
\cos d_{fs}(\rho, \hat{\rho} ) = \sqrt{\tr \rho  \hat{\rho}}~,~
0 \le d_{fs}(\rho, \hat{\rho} ) \le \frac{\pi}{2} .
\end{eqnarray}
the Bures distance $d_{b}$ is defined in the usual way:
\begin{eqnarray}
d_b (\rho, \hat{\rho} ) := \sqrt{ 1- \tr \rho  \hat{\rho}} .
\end{eqnarray}
It is introduced by Bures \cite{bures} in a mathematical context.
\end{defi}
Let $W(\rho,\hat{\rho})$ be a measure of deviation of the measured value 
$\hat{\rho}$ from the actual value $\rho$,
then we have the following equivalent conditions:
\begin{eqnarray*}
 &\circ& W(\rho, \hat{\rho})= W(g\rho g^* , g\hat{\rho}g^*) \hbox{ for }
 g \in \SU (k) , \rho, \hat{\rho} \in \pro . \label{sym} \\
 &\circ& \hbox{There exists a function }h \hbox{ on } [0,1] \hbox{ such that }
 W(\rho ,\hat{\rho}) = h \circ d_{fs}(\rho ,\hat{\rho}) .
\end{eqnarray*}
It is natural to assume that a deviation measure $W(\rho, \hat{\rho})$ is 
monotone increasing with respect to the Fubini-Study distance $d_{fs}$. 

If ${\cal H}_1 , \ldots , {\cal H}_n$ are $n$ Hilbert spaces which 
correspond to the physical systems,
then their composite system is represented by the tensor Hilbert space:
\begin{eqnarray*}
{\cal H}^{(n)} := {\cal H}_1 \otimes \cdots \otimes {\cal H}_n
= \Otimes_{i=1}^{n} {\cal H}_i .
\end{eqnarray*}
Thus, a state on the composite system is denoted by a density operator $\rho$
on ${\cal H}^{(n)}$.
In particular if $n$ element systems $\{ {\cal H}_i \}$ 
of the composite system ${\cal H}^{(n)}$ are independent of each other,
there exists a density $\rho_i$ on ${\cal H}_i$ such that
\begin{eqnarray*}
\rho^{(n)}= \rho_1 \otimes  \cdots \otimes \rho_n = \Otimes_{i=1}^{n} \rho_i .
\end{eqnarray*}
 The condition:
\begin{eqnarray}
{\cal H}_1= \cdots ={\cal H}_n = {\cal H} ,\quad 
\rho_1 = \cdots = \rho_n = \rho \label{iid4}
\end{eqnarray}
corresponds to the independent and identically distributed condition 
(i.i.d. condition) in the classical case.
In this paper, we treat with this estimation problem 
under this condition (\ref{iid4}) called the quantum i.i.d. condition.
This condition means that identical $n$ samples are independently prepared.
The model $\{ \rho^{(n)}= \underbrace{\rho \otimes \cdots \otimes \rho}_{n} 
|\rho  \in \pro \}$ is called $n$-i.i.d. model.  
As $\rho$ is a pure state, ${\cal H}^{(n)}$ and $\rho^{(n)}$ are 
simplified as follows.
Letting $\rho = | \phi \rangle \langle \phi | \in \pro$,
we have 
\begin{eqnarray*}
\rho^{(n)} = \left| \phi^{(n)} \right\rangle \left\langle \phi^{(n)} \right|
~,~ \phi^{(n)} := \overbrace{\phi \otimes \cdots \otimes \phi}^{n}.
\end{eqnarray*}
Because all of the vectors $\phi^{(n)}$ is included in $n$-times 
symmetric tensor space,
for any measurement $M \in {\cal M}(\Omega,{\cal H}^{(n)})$ on the
$n$-times tensor space ${\cal H}^{(n)}$,
the measurement $\tilde{M}(\,d \omega):=  P_{{\cal H}^{(n)}_s}
M (\,d \omega) P_{{\cal H}^{(n)}_s}  \in {\cal M} (\Omega,{\cal H}^{(n)}_s)$
on the $n$-times symmetric tensor space ${\cal H}^{(n)}_s$ satisfies that:
\begin{eqnarray*}
\tr M(\,d \omega) \rho^{(n)} = \tr \tilde{M}(\,d \omega) \rho^{(n)}
\hbox{ for any } \rho \in {\cal H},
\end{eqnarray*}
where ${\cal H}^{(n)}_s$ denotes the $n$-times symmetric tensor space 
on ${\cal H}$.
Therefore, all of possible measurements 
can be regarded as elements of ${\cal M}(\pro,{\cal H}^{(n)}_s)$.
The mean error of the measurement $\Pi \in  {\cal M}({\cal P}({\cal H})
,{\cal H}_s^{(n)})$ with respect to a deviation measure $W(\rho,\hat{\rho})$,
provided that the actual state is $\rho$, is equal to
\begin{eqnarray*}
{\cal D}^{W,(n)}_{\rho}(\Pi)
:= \int_{\pro} W(\rho,\hat{\rho}) \tr ( \Pi (\,d \hat{\rho}) \rho^{(n)}).
\end{eqnarray*}
In minimax approach the maximum possible error with respect to a deviation
measure $W(\rho ,\hat{\rho})$
\begin{eqnarray*}
{\cal D}^{W,(n)}(\Pi)
:= \max_{\rho \in \pro} {\cal D}^{W,(n)}_{\rho}(\Pi)
\end{eqnarray*}
is minimized.
\section{Quantum Hunt-Stein theorem}\label{gun1}
In this section, the quantum Hunt-Stein theorem, established by Holevo 
\cite{HolP,Holc}, is summarized.
Let $G$ be a compact transitive Lie group of all transformations on a compact 
parametric 
set $\Theta$, and $\{ V_g \}$ a continuous unitary irreducible representation 
of $G$ in a finite-dimensional Hilbert space ${\cal H}' := {\bf C}^{k'}$,
and $\mu$ a $\sigma$-finite invariant measure on group $G$ such that 
$\mu(G)=1$.
In this section, we consider the following condition for a measurement.
\begin{defi}
A measurement $\Pi \in {\cal M}(\Theta,{\cal H}')$ is covariant with 
respect to $\{ V_g \}$ if 
\begin{eqnarray*}
 V_g^*  \Pi ( B) V_g
 = \Pi ( B_{g^{-1}}) 
\end{eqnarray*}
for any  $g \in G$ and any Borel $B \subset \Theta$,
 where
\begin{eqnarray*}
B_{g} := \{ g \theta | \theta \in B \}.
\end{eqnarray*}
${\cal M}(\Theta,V)$ denotes the set of covariant measurements with respect 
to $\{ V_g \}$.
\end{defi}
Covariant measurements are characterized by the following theorem.
\begin{thm}\label{kyo1}
The map $V^{\theta}$ from the set ${\cal S}({\cal H}')$ of densities
on ${\cal H}'$ to ${\cal M}(\Theta,V)$ is surjective for any $\theta \in
\Theta$, where $V^{\theta}(P)$ is defined as follows:
\begin{eqnarray*}
 V^{\theta}(P) (B) :=
 k' \int_{\{ g \theta \in B \} } V_g P V_g^* \mu (\,d g)
\hbox{ for } B \in {\cal B}(\Theta) 
\end{eqnarray*}
for any $P \in {\cal S}({\cal H}')$.
\end{thm}
In this section, we treat with the following condition for a family of states.
\begin{defi}
The family is called covariant under the representation $\{ V_g \}$ of 
group $G$ acting on $\Theta$, if 
\begin{eqnarray*}
S_{ g \theta} = V_g S_{ \theta} V_g^*, \quad \forall g \in G, 
\forall \theta \in \Theta .
\end{eqnarray*}
\end{defi}
 Assuming that the object is prepared in one of the states $\{ S_{\theta} |
\theta \in \Theta \}$ but
the actual value of $\theta$ is unknown, then the difficulty
is estimating this 
value as close as possible to a measurement on the object.
We shall solve this problem by means of the quantum statistical decision 
theory.

Let $W(\theta,\hat{\theta})$ be a measure of deviation of the measured value
$\hat{\theta}$ from the actual value $\theta$.
It is natural to assume that $W(\theta,\hat{\theta})$ is invariant:
\begin{eqnarray}
W( \theta, \hat{\theta})= W( g \theta , g \hat{ \theta} ) \quad \hbox{ for }
\forall g \in G , \forall \theta, \forall \hat{\theta} \in \Theta \label{sym1}.
\end{eqnarray}
The mean error of the measurement $ \Pi \in  {\cal M}(\Theta,{\cal H}')$ with 
respect to a deviation measure $W(\theta ,\hat{\theta})$,
provided that the actual state is $S_{\theta}$, is equal to
\begin{eqnarray*}
{\cal D}^{W,S}_{\theta}(\Pi)
:= \int_{\Theta} W(\theta ,\hat{\theta}) \tr ( \Pi (\,d \hat{\theta}) S_{\theta}) .
\end{eqnarray*}
Following the classical statistical decision theory,
we can form two functionals of ${\cal D}^{W}_{\theta}$ 
giving a total measure of precision of the measurement $\Pi$.

In Bayes' approach we take the mean of ${\cal D}^{W}_{\theta}$ with respect to
a given prior distribution $\pi (\,d \theta)$. The measurement minimizing 
the resulting functional:
\begin{eqnarray*}
{\cal D}_{\pi}^{W,S}(\Pi) 
:= \int_{\Theta}{\cal D}_{\theta}^{W,S} (\Pi) \pi(\,d \theta)
\end{eqnarray*}
is called Bayesian.
This quantity represents the mean error in the situation where $\theta$ is a
random parameter with known distribution $\pi (\,d \theta)$. 
In particular, as $\Theta,G$ are compact and ``nothing is known'' about 
$\theta$, it is natural to take for $\pi (\,d \theta)$ the ``uniform'' 
distribution, i.e. normalized invariant measure $\nu (\,d \theta)$ 
defined as follows:
\begin{eqnarray*}
\nu( B ) := \mu (\{ g \theta \in B \} ).
\end{eqnarray*}
It is independent of the choice of $\theta \in \Theta$.

In minimax approach the maximum possible error with respect to a deviation 
measure $W(\theta,\hat{\theta})$
\begin{eqnarray*}
{\cal D}^{W,S}(\Pi)
:= \max_{\theta \in \Theta} {\cal D}^{W,S}_{\theta}(\Pi)
\end{eqnarray*}
is minimized.
The minimizing measurement is called minimax.

Because $G$ is compact, we shall show that in the covariant case the minima of
Bayes and minimax criteria coincide and are achieved on a covariant
measurement.
We obtain the following quantum Hunt-Stein theorem \cite{HolP,Holc}.
It is easy to prove the theorem.
\begin{thm}
For a covariant measurement $\Pi \in {\cal M}(\Theta,V)$,
we obtain the following equations:
\begin{eqnarray*}
{\cal D}_{\theta}^{W,S}(\Pi)
= {\cal D}_{\nu}^{W,S}(\Pi)
={\cal D}^{W,S}(\Pi) .
\end{eqnarray*}
\end{thm}
For $\Pi \in {\cal M}(\Theta,{\cal H}')$,
denote 
\begin{eqnarray*}
\Pi_g (B) :=
V_g \Pi (B_g) V_g^* \quad \hbox{ for } B \in {\cal B}(\Theta).
\end{eqnarray*}
Introducing the ``averaged'' measurement 
\begin{eqnarray*}
\bar{\Pi} (B) :=
\int_{G} \Pi_{g^{-1}} (B) \mu( \,d g) ,
\end{eqnarray*}
we have
\begin{eqnarray*}
{\cal D}_{\nu}^{W,S}(\bar{\Pi})
= \int_{G} {\cal D}_{\nu}^{W,S}(\Pi_{g^{-1}}) \mu(\,d g)
={\cal D}_{\nu}^{W,S}(\Pi) .
\end{eqnarray*}
Thus,
\begin{eqnarray*}
{\cal D}^{W,S}(\Pi) \ge {\cal D}_{\nu}^{W,S}(\Pi)
= {\cal D}_{\nu}^{W,S}(\bar{\Pi}).
\end{eqnarray*}
In this case, minimax approach and Bayes' approach with respect to 
$\nu ( \,d \theta)$ are equivalent.
Therefore we minimize the following:
\begin{eqnarray*}
{\cal D}_{\theta}^{W,S} \circ V^{\theta} (P)
= k' \int_G W(\theta, g \theta) \tr S_{\theta} V_g P V_g^* \mu (\,d g)
= \tr \hat{W}(\theta) P ,
\end{eqnarray*}
where
\begin{eqnarray*}
\hat{W}(\theta)  &:=&
 k' \int_G W(\theta, g \theta) V_g^*  S_{\theta} V_g  \mu (\,d g) \\
 &=&
 k' \int_{\Theta} W(\theta, \hat{\theta} ) S_{\hat{\theta}} 
\nu (\,d \hat{\theta}) . 
\end{eqnarray*}
Thus, it is sufficient to consider the following minimization:
\begin{eqnarray*}
\min_{ P \in {\cal S}({\cal H})} \tr \hat{W}(\theta)  P
=\min_{ P \in \proa } \tr \hat{W}(\theta)  P .
\end{eqnarray*}

\section{Optimal measurement in pure state $n$-i.i.d. model}\label{saiteki}
In this section we apply the theory of \S \ref{gun1} to the problem 
\S \ref{pure1}.

We let as follows:
\begin{eqnarray*}
\Theta := \pro ,~ 
{\cal H}' := {\cal H}^{(n)}_s ,~
G := \SU (k) ,~
S_{ \rho} := \rho^{(n)}.
\end{eqnarray*}
Then, the invariant measure $\nu$ on $\pro$ is equivalent to the measure 
defined by the volume bundle induced by the Fubini-Study metric.
We let the action $\{ V_g \}$ of $G=\SU (k)$ to ${\cal H}^{(n)}_s $ be 
 the tensor representation of the natural representation.
 In this case, we have $k' = {n+k-1 \choose k-1}$.
\begin{thm}\label{main}
If a deviation measure $W(\rho,\hat{\rho})$ is 
monotone increasing with respect to the Fubini-Study distance $d_{fs}$,
then we get 
\begin{eqnarray*}
 \min_{ P_0 \in {\cal P}({\cal H}^{(n)}_s) } \tr \hat{W}(\rho) P_0
 = \tr \hat{W}(\rho) \rho^{(n)} .
\end{eqnarray*}
\end{thm}
For a proof see Appendix \ref{proof}. 
Thus, $V^{\rho}(\rho^{(n)})$ is the optimal measurement with respect to 
a deviation measure $W(\rho,\hat{\rho})$.
The optimal measurement is independent of the choice of $\rho$ and $W$ 
since $V^{\rho_0}(\rho_0^{(n)})=V^{\rho}(\rho^{(n)})$.  
This optimal measurement is denoted by $\Pi_n$ and is described as follows:
\begin{eqnarray*}
\Pi_n (\,d \hat{\rho} ) := {n+k-1 \choose k-1} \hat{\rho}^{(n)} \nu 
(\,d \hat{\rho}) .
\end{eqnarray*}
Under the following chart (\ref{tei5}), 
the optimal measurements are denoted as:
\begin{eqnarray}
\Pi_n (\,d \theta) =
{n+k-1 \choose k-1} \left | \phi (\theta) ^{(n)} \right \rangle 
\left \langle \phi (\theta)^{(n)} \right |  \nu (\,d \theta) \label{alpha}
\end{eqnarray}
for 
$\theta \in \{ \theta \in {\bf R}^{2k-2}| \theta_i \in [0, 2 \pi ) 
1 \le j \le k-1 , \theta_j \in [0, \pi / 2 ] \} $,
where we defined as follows:
\begin{eqnarray}
\phi(\theta) 
&:=&
\begin{pmatrix}
\cos \theta_1 \\
e^{i \theta_{k}} \sin \theta_1 \cos \theta_2 \\
e^{i \theta_{k+1}} \sin \theta_1 \sin  \theta_2 \cos \theta_3 \\
\vdots \\
e^{i \theta_{2k-3}} \sin \theta_1 \sin  \theta_2 \sin \theta_3 
\cdots \sin \theta_{k-2} \cos \theta_{k-1} \\
 e^{i \theta_{2k-2}} \sin \theta_1 \sin  \theta_2 \sin \theta_3 
\cdots \sin \theta_{k-2} \sin \theta_{k-1} \\
\end{pmatrix} \label{tei5}.
 \end{eqnarray}
 The invariant measures $\nu (\,d \theta)$ described above is from 
[21,p.31].
 \begin{eqnarray}
 \nu (\,d \theta) = \frac{(k-1) !}{ \pi^{k-1}}
 \sin^{2k-3} \theta_1 \sin^{2k-5} \theta_2 \cdots \sin \theta_{k-1} 
 \cos \theta_1 \cos \theta_2 \cdots \cos \theta_{k-1} 
 \,d \theta_1 \,d \theta_2 \cdots \,d \theta_{2k-2}. \label{beta}
 \end{eqnarray}
 \begin{lem}\label{bunpu}
 If the deviation measure $W$ is characterized as $W(\rho,\hat{\rho})= 
h \circ d_{fs} (\rho, \hat{\rho})$, we can describe the maximum possible 
error of the optical measurement $\Pi_n$ as:
 \begin{eqnarray*}
 {\cal D}^{W,(n)}(\Pi_n)
 = 2 (k-1) {n+k-1 \choose k-1} \int_0^{\frac{\pi}{2}} h( \theta ) 
\cos^{2n+1} \theta 
 \sin^{2k-3} \theta \,d \theta .
 \end{eqnarray*}
 \end{lem}
 For a proof, see Appendix \ref{pf1}.
 
Next, we asymptotically calculate the error of the optimal measurements 
$\Pi_n$ in the third order.
 \begin{thm}\label{th61}
When the deviation measure $W$ is described as $W = d_b^{\gamma}$,
we can asymptotically
calculate the maximum possible error of the optimal measurement as:
 \begin{eqnarray}
 \lim_{n \to \infty} {\cal D}^{d_b^{\gamma},(n)}(\Pi_n) n^{\frac{\gamma}{2}}
 =\frac{\Gamma( k-1+\gamma/2)}{\Gamma( k-1)}. \label{kere} 
 \end{eqnarray}
 Specially in the case of $\gamma =2$, we have
 \begin{eqnarray}
  {\cal D}^{d_b^{2},(n)}(\Pi_n) n
 =\frac{(k-1)n}{n+k} 
= (k-1)\sum_{i=0}^{\infty}\left(-\frac{k}{n}\right)^{i}
\to k-1 \hbox{ as } n \to \infty . \label{keha}
 \end{eqnarray}
When the deviation measure is defined by the square of the Fubini-Study 
distance,
we can asymptotically calculate the maximum possible error of the optimal 
measurement as:
 \begin{eqnarray}
{\cal D}^{d_{fs}^2,(n)}(\Pi_n) n
\cong
(k-1) - \frac{2}{3}k(k-1)\frac{1}{n}
+ k(k-1)\frac{23k-7}{45} \frac{1}{n^2}
\hbox{ as } n \to \infty . \label{bun1}
\end{eqnarray}
The error of the sequence $\{ \Pi_n \}_{n=1}^{\infty}$
of the optimal measurements 
can be calculated in the sense of large deviation as:
\begin{eqnarray}
 &~& \frac{1}{n}\log \left(
 \Pr_{\Pi_n}^{\rho^{(n)}} \{ \hat{\rho} \in \pro | d_{fs}(\rho, \hat{\rho}) 
\ge \epsilon \}  \right) \nonumber\\
&\cong& \log \cos^2 \epsilon +(k-2) \frac{\log n}{n}
+ \left(  - \log (k-2) ! + 2(k-2) \log (\sin \epsilon )- 2 \log 
(\cos \epsilon) 
\right) \frac{1}{n} \nonumber\\
&~&+  \left(
\frac{ k ^2 -k -2 }{2} + (k-2)\cot^2 \epsilon \right) \frac{1}{n^2}
\hbox{ as } n \to \infty , \label{zenkin3}
\end{eqnarray}
where
 $\Pr_{M}^{S}B$ denotes the probability of $B$
 with respect to the probability measure $\tr (M(\,d \omega) S )$ 
 for a Borel $B \subset \Omega$, a measurement $M \in {\cal M}
(\Omega,{\cal H}')$ and a state $S \in {\cal S}({\cal H}')$.
\end{thm}
For a proof, see Appendix \ref{pf2}.
The first term of the right hand of (\ref{zenkin3}) coincide with
the logarithm of the fidelity.
About the fidelity, see Jozsa \cite{jozsa}.

In this paper, $\epsilon$ in equations (\ref{zenkin3}) is called admissible
radius.

Since
\begin{eqnarray}
 \lim_{\epsilon \to 0}\frac{\log \cos^2 \epsilon}{\epsilon^2} 
 = - 1 , \label{hahu}
\end{eqnarray}
 we obtain the following large deviation approximation.
\begin{eqnarray}
 \lim_{\epsilon \to 0} \lim_{n \to \infty}  
 \frac{1}{\epsilon^2 n} \log \left(
 \Pr_{\Pi_n}^{\rho^{(n)}} \{ \hat{\rho} \in \pro | d_{fs}(\rho, \hat{\rho}) 
\ge \epsilon \} \right)
= -1 . \label{zenkin5} 
\end{eqnarray}
 
\section{Semiclassical measurement}\label{fish}
In this section, we consider measurements allowed the 
quantum correlation between finite samples only.
A measurement $M$ on ${\cal H}^{(nm)}$ is called {\it $m$-semiclassical} 
if there exists a estimator $T$ on the probability space 
$\underbrace{\pro \times \cdots \times \pro}_{n}$ 
whose range is $\pro$ such that 
\begin{eqnarray}
M(B)= \int_{T^{-1}(B)} \underbrace{\Pi_m (\,d \rho_1 )\otimes \cdots 
\otimes \Pi_m(\,d \rho_n)}_{n} ~\forall B \subset \pro . \label{semic}
\end{eqnarray}
We compare the error between $m$-semiclassical measurements 
and the optimal measurement $\Pi_{nm}$ for $nm$ samples of the unknown state
as the equations (\ref{bun1}),(\ref{zenkin3}),(\ref{zenkin5}).

In doing this comparison, we bear in mind 
asymptotic estimation theory in classical statics.
In classical statics, it is assumed that the sequence of estimators satisfies
the consistency.
\begin{defi}
A sequence  $\{ T^{(n)} \}_{n=1}^{\infty}$ of estimators on a 
probability space $\Omega$ is called consistent
with respect to a family $\{ p_{\theta} | \theta \in \Theta \}$ of 
probability distributions on $\Omega$,
if it satisfies the condition (\ref{hama}),
where every $T^{(n)}$ is a probability variable on  
the probability space 
$\underbrace{\Omega \times \cdots \times \Omega}_{n}$ 
whose range is $\Theta$.
\begin{eqnarray}
p^{(n)}_{\theta} \{d_J(T^{(n)},\theta) \,> \epsilon \} \to 0
\hbox{ as } n \to \infty
, \forall \theta \in \Theta, \forall  \epsilon \,> 0 , \label{hama}
\end{eqnarray}
where $d_J$ denotes the geodesic distance defined by the Fisher Information
metric and 
$p^{(n)}_\theta$ denotes the probability measure 
$\underbrace{p_{\theta} \times \cdots \times p_{\theta}}_{n}$
on the probability space $\underbrace{\Omega \times \cdots \times \Omega}_{n}$.
\end{defi}
It is well known that the following theorem establishes
under the preceding consistency \cite{Ba1,Fu1,Fu2}. 
\begin{thm}\label{app}
If a sequence $\{ T^{(n)} \}_{n=1}^{\infty}$ of estimators 
is a consistent estimator with respect to 
 a family $\{ p_{\theta} | \theta \in \Theta \}$ of 
probability distributions on a probability space $\Omega$
which satisfies some regularity, then we have the following inequalities.
\begin{eqnarray}
\lim_{n \to \infty} n \underbrace{\int \int \cdots \int}_{n}
d_{J}^2 \left( T^{(n)}(x_1, x_2, \ldots , x_n ) , \theta \right) 
p^{(n)}_{\theta}(\,d x_1, \ldots , \,d x_n) 
&\ge& \dim \Theta \label{bun} \\
\lim_{n \to \infty} \frac{1}{n}
\log \left( p^{(n)}_{\theta} \{ D( p_{T^{(n)}} \| p_{\theta} ) \ge \epsilon \}
 \right) &\ge& -\epsilon \label{div} \\
\lim_{\epsilon \to 0} \lim_{n \to \infty} \frac{1}{\epsilon^2 n}
\log \left (p^{(n)}_{\theta} \{ d_J( T^{(n)}, \theta ) \ge \epsilon \} 
\right) &\ge& -\frac{1}{2} , \label{fi}
\end{eqnarray}
where $D(p \| q)$ denotes the information divergence of a probability 
distribution $q$ with respect to another probability 
distribution $p$ defined by:
\begin{eqnarray*}
D(p \| q) := \int_{\Omega}\left( \log p(\omega) - \log q(\omega) \right)
p(\omega) \,d \omega.
\end{eqnarray*}
Under some regularity conditions, the lower bounds of (\ref{bun}),(\ref{fi}) 
can be attained by the maximum likelihood estimator(MLE).
A regularity condition for attaining (\ref{bun}) is different from
a one for (\ref{fi}).
The lower bound of (\ref{div}) can be attained by
the MLE when the family $\{ p_{\theta} | \theta \in 
\Theta \}$ is exponential, but generally cannot be attained.
\end{thm}
For the comparison,
we apply Theorem \ref{app} to the family of distributions
$\{\tr \Pi_m( \,d \hat{\rho} ) \rho^{(m)} | \rho \in \pro \}$
given by the measurement $\Pi_m$ and 
the family of states $\{ \rho^{(m)} | \rho \in \pro \}$.
Let $T_{(n,m)}$ be the  measurement on ${\cal H}^{(nm)}$
defined by the estimator $T^{(n)}$ and $n$ data 
given by the measurement $\underbrace{\Pi_m \otimes \cdots \otimes \Pi_m}_{n}$ 
and the state $\rho^{(nm)}$.
We consider the sequence of measurements $\{ T_{(n,m)} \}_{n=1}^{\infty}$.
From the symmetry of $\pro$ and $\Pi_m$, the information divergence of 
a probability measure $\tr \Pi_m ( \,d \hat{\rho})\rho_1^{(m)}$ 
with respect to another a probability measure $\tr \Pi_m ( \,d \hat{\rho})
\rho_2^{(m)}$ is determined by the the Fubini-Study distance
$\epsilon$ between $\rho_1$ and $\rho_2$.
Thus, the divergence can be denoted by $D_m(\epsilon)$.
From Lemma \ref{kaka}, the geodesic distance 
$d_{\Pi_m}$ with respect to Fisher information metric in
the family of distributions 
 $\{\tr \Pi_m( \,d \hat{\rho} ) \rho^{(m)} | \rho \in \pro \}$
is given by:
\begin{eqnarray*}
d_{\Pi_m} = \sqrt{2 m} d_{fs}.
\end{eqnarray*}
Since $\dim \pro = 2(k-1)$, we have the following inequalities:
\begin{eqnarray}
\lim_{n \to \infty} nm {\cal D}^{d_{fs}^2,(nm)}(T_{(n,m)}) 
= \lim_{n \to \infty} \max_{\rho \in \pro} 
nm \int_{\pro} d^2_{fs}(\rho,\hat{\rho}) 
\tr (T_{(n,m)} (\,d \hat{\rho}) \rho^{(nm)} )
&\ge& k-1 \label{bun2}\\
\lim_{n \to \infty} \frac{1}{nm}
\log \Pr_{T_{(n,m)}}^{\rho^{(nm)}} \{ \hat{\rho} \in \pro | d_{fs}
(\rho, \hat{\rho}) \ge \epsilon \} 
&\ge& -\frac{D_m(\epsilon)}{m} \label{div2} \\
\lim_{\epsilon \to 0} \lim_{n \to \infty}  
\frac{1}{\epsilon^2 nm} \log 
\Pr_{T_{(n,m)}}^{\rho^{(nm)}} \{ \hat{\rho} \in \pro | d_{fs}
(\rho, \hat{\rho}) \ge \epsilon \} 
& \ge& -1 .\label{fi2}
\end{eqnarray}
The lower bound of (\ref{bun2}) is consistent with 
the first term of the right hand of (\ref{bun1}) and 
the lower bound of (\ref{fi2}) is consistent with 
the right hand of (\ref{zenkin5}).
The family of distributions 
 $\{\tr \Pi_m( \,d \hat{\rho} ) \rho^{(m)} | \rho \in \pro \}$
satisfies a regularity condition for (\ref{bun}) by MLE.
But, we cannot show that it does a one for (\ref{fi}).
We have the following lemma
concerning the comparison of the lower bound
$-\frac{D_m(\epsilon)}{m}$
of (\ref{div2}) and the first term
$2\log \cos \epsilon$ of the right hand of (\ref{zenkin3}).
\begin{lem}\label{kaka}
We can calculate the divergence $D_m(\epsilon)$ and 
the distance $d_{\Pi_m}$ as:
\begin{eqnarray}
\frac{D_m(\epsilon)}{m} 
&=& \sum_{i=1}^{m}\frac{\sin^{2i} \epsilon}{i}
\to -\log \left( 1- \sin^2 \epsilon \right)
= - \log \cos^2 \epsilon \quad \hbox{ as } m \to \infty  \label{kiuu} \\
d_{\Pi_m} 
&=& \sqrt{2m}d_{fs} \label{kaka1}.
\end{eqnarray}
Therefore, 
$\frac{D_m(\epsilon)}{m}$ is monotone increasing with respect to $m$.
\end{lem}
For a proof, see Appendix \ref{pf3}.
(\ref{kiuu}) derives that
\begin{eqnarray}
0 \,< 
\frac{- m \log \cos^2 \epsilon -D_m(\epsilon)}{m \epsilon^{2m}} \to 0
\hbox{ as } \epsilon \to 0
\label{un}
\end{eqnarray}
which means that 
the first term of (\ref{zenkin3}) cannot be
attained by a semi-classical measurement.
However,
it is an open problem as to whether 
the left-hand side of (\ref{zenkin5})
can be asymptotically attained by a $1$-semiclassical measurement.
Concerning MSE, the first term of (\ref{bun1}) 
can be asymptotically attained by it i.e. 
it can be asymptotically attained by measurements without 
using quantum correlations between samples.
Thus, in order to attain it asymptotically,
it is sufficient 
to physically realize the optimal measurement $\Pi_1$
on a single sample.
Indeed, $\Pi_1$ is a random measurement as follows.
To denote $\Pi_1$ as a random measurement,
we will define the simple measurement $E_g (g \in \SU (k))$
whose measurable space $\pro$.
For an element $g \in \SU (k)$, the vectors $\phi_1(g), \cdots , \phi_k(g)$
in ${\cal H}$ are defined as:
\begin{eqnarray*}
\left( \phi_1(g) \cdots \phi_k(g) \right) = g.
\end{eqnarray*}
The measurement $E_g$ is defined as:
\begin{eqnarray*}
E_g\left( | \phi_i(g) \rangle \langle \phi_i(g) | \right) = 
 | \phi_i(g) \rangle \langle \phi_i(g) |.
\end{eqnarray*}
Therefore, the optimal measurement $\Pi_1$ for a single sample
can be described as the following random measurement:
\begin{eqnarray}
\Pi_1 =
\int_{\SU (k)} E_g \mu (\,d g ), \label{hor}
\end{eqnarray}
where $\mu$ is the invariant measure on $\SU (k)$ with $\mu (\SU (k))=1$.
Therefore, in order to realize the optimal measurement $\Pi_1$,
it is sufficient to realize the simple measurement
$E_g$ for any $g \in \SU (k)$.

\section{Conclusion}
We have compared two cases.
One regards the system consisting of enough samples 
as the single system,
the other regards it as separate systems.
Under this comparison,
the MSEs of both cases asymptotically coincide in the first order
with respect to the Fubini-Study distance
(see (\ref{bun1}) and (\ref{bun2})). 
However we leave the question of whether 
they asymptotically coincide in the second order
with respect to the Fubini-Study distance to a future study. 
On the other hand, in view of the evaluation of large deviation,
if the allowable radius is finite,
neither coincide (see (\ref{zenkin3}) and (\ref{div2})).
However, 
in the case of the allowable radius goes to infinitesimal,
it is an open problem as to whether both coincide (see (\ref{zenkin5}) 
and (\ref{fi2})).

These results depend on the effect of a pure state.
Therefore, it is an open question as to whether
the MSEs of both cases asymptotically coincide in the first order
in another family.
In the case of large deviation, the same question is also open
in the limit where the radius of allowing error
goes to infinitesimal.

\section*{Acknowlegments}
The author wishes to thank Prof. K. Ueno.
Also, he wishes to thank Dr. A. Fujiwara, Dr. K. Matsumoto and Dr. H. Nagaoka.
\par\bigskip
\indent
{\LARGE\bf Appendices}
\appendix
\section{Proof of Theorem 3}\label{proof}
In this Appendix, assume that $\rho = | \phi (0) \rangle \langle \phi (0) |$.
Because ${\cal H}^{(n)}_s$ is irreducible with respect to the action of
 $\SU (k)$,
\begin{eqnarray}
{\cal H}^{(n)}_s
&=& \left\{\left. \sum_{i} a_i V_{g_i} \phi(0)^{(n)} \right| a_i \in {\bf C}, g_i \in \SU (k) \right\} \nonumber \\
&=& \left\{\left. \sum_{i} \phi_i^{(n)} \right| 
\phi_i \in {\cal H} \right\}. \label{kiya}
\end{eqnarray}
We assume that $W( \rho, \hat{\rho} ) = h (\tr \rho \hat{\rho})$.
As $h$ is monotone decreasing, 
there exists a measure $h'$ on $[0,1]$ such that
$h( x ) = h'( [ x, 1])$.

The function $h_{\beta}$ on $[0,1]$ and the deviation measure $W_{\beta}$
are defined as follows:
\begin{eqnarray*}
h_{\beta}(x) &:=&
\left \{
\begin{matrix}
1 & \hbox{for} & x \le \beta \\
0 & \hbox{for} & x \,>  \beta  
\end{matrix}
\right . \\
W_{\beta}(\rho,\hat{\rho})& :=& h_{\beta}(\tr \rho \hat{\rho} ).
\end{eqnarray*}
From Lemma \ref{le1}, for any measurement $\Pi$ we have
\begin{eqnarray*}
{\cal D}_{\rho}^{W,(n)}(\Pi)
= \int_{ [0,1] } {\cal D}_{\rho}^{W_{\beta},(n)}( \Pi) h' (\,d \beta) .
\end{eqnarray*}
From (\ref{kiya}), it is sufficient to show 
the following for $\{ \phi_i \} \subset {\cal H}$
in the case of $W=W_{\beta}$.
\begin{eqnarray}
\frac{\tr \hat{W_{\beta}}(\rho) \left| \sum_{i} \phi_i^{(n)} \right\rangle 
\left\langle  \sum_{i} \phi_i^{(n)} \right|}
{ \left\langle  \sum_{i} \phi_i^{(n)} \left| \sum_{i} \phi_i^{(n)} \right.
\right\rangle }
\ge 
\tr \hat{W_{\beta}}(\rho) \left| \phi(0)^{(n)} \right\rangle 
\left\langle \phi(0)^{(n)} \right| .
\label{shou}
\end{eqnarray}
From Lemma \ref{le2} it is sufficient for (\ref{shou}) to prove the following:
\begin{eqnarray}
&~& \left\langle  \sum_{i} \phi_i^{(n)} \right | \hat{W_{\beta}}(\rho) 
\left| \sum_{i} \phi_i^{(n)} \right\rangle \cdot
\left\langle \phi(0)^{(n)} \right| \id -  \hat{W_{\beta}}(\rho) 
\left| \phi(0)^{(n)} \right\rangle 
\nonumber \\
&\ge&
\left\langle \phi(0)^{(n)} \right|\hat{W_{\beta}}(\rho) 
\left| \phi(0)^{(n)} \right\rangle \cdot 
\left\langle  \sum_{i} \phi_i^{(n)} \right| \id -  \hat{W_{\beta}}(\rho)  
\left|\sum_{i} \phi_i^{(n)} \right\rangle 
\label{shou1} .
\end{eqnarray}
Remark that $| \langle \phi(\theta) | \phi (0) \rangle |^2 = \cos^2 \theta_1$.
From Lemma \ref{le3}, we get 
\begin{eqnarray*}
\left\langle  \sum_{i} \phi_i^{(n)} \right| \hat{W_{\beta}}(\rho) 
\left| \sum_{i} \phi_i^{(n)} \right\rangle 
&=& \frac{k' \cdot (k-1)!}{\pi^{(k-1)}}
\int_{\alpha}^{\frac{\pi}{ 2}} f_1( \theta_1) \cos \theta_1 \sin^{2k-3} 
\theta_1 \,d \theta_1 \\
\left\langle  \sum_{i} \phi_i^{(n)} \right| \id - \hat{W_{\beta}}(\rho) 
\left| \sum_{i} \phi_i^{(n)} \right\rangle 
&=& \frac{k' \cdot (k-1)!}{\pi^{(k-1)}}
\int_0^{\alpha} f_1( \theta_1)  \cos \theta_1 \sin^{2k-3} \theta_1\,d 
\theta_1 \\
\left \langle  \phi(0)^{(n)} \right| \hat{W_{\beta}}(\rho) 
\left| \phi(0)^{(n)} \right\rangle
&=& C \int_{\alpha}^{\frac{ \pi}{2}} 
\cos^{2n+1} \theta_1 \sin^{2k-3} \theta_1
 \,d \theta_1  \\
\left \langle  \phi(0)^{(n)} \right| \id - \hat{W_{\beta}}(\rho) 
\left| \phi(0)^{(n)} \right\rangle 
&=& C \int_0^{\alpha} 
\cos^{2n+1} \theta_1 \sin^{2k-3} \theta_1
 \,d \theta_1 ,
\end{eqnarray*}
where 
\begin{eqnarray*}
 \beta &:=& \cos^2 \alpha \\
f_1(\theta_1) 
&:=& 
 \underbrace{\int_0^{\frac{\pi}{2}} \cdots \int_0^{\frac{\pi}{2}}}_{k-2}
f_2(\theta_1, \ldots , \theta_{k-1} )
\lambda( \,d \theta_2 \cdots \,d \theta_{k-1}) \\
f_2(\theta_1, \ldots , \theta_{k-1}) 
&:=& \underbrace{\int_0^{2 \pi} \cdots \int_0^{2 \pi}}_{k-1}  
\sum_{i,j}
\langle \phi_i | \phi(\theta) \rangle ^n
\langle \phi ( \theta) | \phi_j \rangle ^{n}  
\,d \theta_{k} \cdots \,d \theta_{2k-2} \\ 
C
&:=& \frac{k' \cdot (k-1)!}{\pi^{(k-1)}} 
\underbrace{\int_0^{2 \pi} \cdots \int_0^{2 \pi}}_{k-1}  
\underbrace{\int_0^{\frac{\pi}{2}} \cdots \int_0^{\frac{\pi}{2}}}_{k-2}
\lambda( \,d \theta_2 \,d \theta_3 \cdots \,d \theta_{k-1})
\,d \theta_{k} \cdots \,d \theta_{2k-2}  \\
\lambda( \,d \theta_2 , \ldots , \,d \theta_{k-1})
&:=& \sin^{2k-5} \theta_2 
\cdots \sin \theta_{k-1} 
\cos \theta_2 \cdots \cos \theta_{k-1} 
\,d \theta_2 \cdots \,d \theta_{k-1} .
\end{eqnarray*}
Therefore, it is sufficient for the equation (\ref{shou1}) to show 
that for $\pi /2 \ge \theta_1 \,> \theta_1' \ge 0 $ 
\begin{eqnarray*}
f_1 (\theta_1) \sin^{2k-3} \theta_1 \cos^{2n+1} \theta_1' 
\sin^{2k-3} \theta_1' \ge f_1(\theta_1') \sin^{2k-3} \theta_1'  
\cos^{2n+1} \theta_1 \sin^{2k-3} \theta_1 .
\end{eqnarray*}
It suffices to verify that for 
$\theta_i \in [0, \frac{\pi}{2}] ,2 \le i \le k-1 ,\pi /2 \ge \theta_1 \,> 
\theta_1' \ge 0$ 
\begin{eqnarray*}
\frac{f_2 (\theta_1 ,\theta_2, \ldots , \theta_{n-1} )}{\cos^{2n} \theta_1 }
 \ge 
\frac{f_2 (\theta_1' ,\theta_2, \ldots , \theta_{n-1} )}{\cos^{2n} \theta_1' }
 .
\end{eqnarray*}
Thus, it is sufficient to prove that 
the following is monotone decreasing about $\theta_1$ 
for any $\theta_2 , \ldots , \theta_{k-1}$:
\begin{eqnarray}
\frac{1}{\cos^{2n} \theta_1 }
 \underbrace{\int_0^{2 \pi} \cdots \int_0^{2 \pi}}_{k-1}  
\sum_{i,j}
\langle \phi_i | \phi(\theta) \rangle ^n
\langle \phi ( \theta ) | \phi_j \rangle ^{n}  
\,d \theta_{k} \cdots \,d \theta_{2k-2} \label{tan1} .
\end{eqnarray}
Letting
\begin{eqnarray*}
\phi_i 
:= 
\begin{pmatrix}
e^{i \psi_i^1}\phi_i^1 \\
e^{i \psi_i^2} \phi_i^2 \\
\vdots \\
e^{i \psi_i^{k}}\phi_i^{k} 
\end{pmatrix}
 ,
\end{eqnarray*}
we get 
\begin{eqnarray*}
&~& \frac{\langle \phi_i | \phi(\theta) \rangle ^n}{ \cos^{n} \theta_1 } \\
& =& \Bigl( 
e^{i \psi_i^1} \phi_i^1 +
\sum_{j=2}^{k-1 } e^{i (\theta_{k-2+j}- \psi_i^j)}
\tan \theta_1 \sin \theta_2 \cdots \sin \theta_{j} \cos \theta_{j+1}
\phi_i^{j} \\
&~& + e^{i (\theta_{2k-2}- \psi_i^{k-1})}
\tan \theta_1 \sin \theta_2 \cdots \sin \theta_{k-1} 
\phi_i^{k}
\Bigr)^n .
\end{eqnarray*}
Letting $x:=\tan \theta_1$,
Lemma \ref{lem1} induce that (\ref{tan1}) is monotone decreasing 
about $\theta_1$.
The proof is complete.
\begin{lem}\label{le1}
If the deviation measure $W( \rho , \hat{\rho} )
= h' ( [ \tr \rho \hat{\rho}  , 1 ])$, then 
\begin{eqnarray}
{\cal D}_{\rho}^{W,(n)}(\Pi)
= \int_{ [0,1] } {\cal D}_{\rho}^{W_{\beta},(n)}( \Pi) h' (\,d \beta) . 
\label{le11}
\end{eqnarray}
\end{lem}
\begin{pf}
For the probability measure $\pi$ on $\pro$,
we have 
\begin{eqnarray*}
\int_{\pro} W( \rho,\hat{ \rho}) \pi (\,d \hat{\rho}) 
&=& \int_{\pro} h( \tr \rho \hat{ \rho} ) \pi ( \,d \hat{\rho}) \\
&=& \int_{\pro} \int_{ [0,1]} h_{\beta} ( \tr \rho \hat{ \rho} ) 
h'(\,d \beta) \pi(\,d \hat{\rho}) \\
&=& \int_{ [0,1]} 
\Bigl( \int_{\pro} h_{\beta} ( \tr \rho \hat{\rho} ) 
\pi(\,d \hat{\rho}) \Bigr)
h'(\,d \beta) \\
&=& \int_{ [0,1]} 
\Bigl( \int_{\pro} W_{\beta}( \rho,\hat{ \rho} ) 
\pi(\,d \hat{\rho}) \Bigr)
h'(\,d \beta) .
\end{eqnarray*}
Substituting $\pi ( \,d \hat{\rho}) $ for 
$\tr ( \Pi(\,d \hat{\rho}) \rho^{(n)})$,
then we obtain (\ref{le11}).
\end{pf}
\begin{lem}\label{le2}
Let ${\cal H}$ be any finite dimensional Hilbert space.
For any elements $\phi,\psi \in {\cal H}$ and any selfadjoint operator $A$ on
${\cal H}$,
the following are equivalent.
\begin{eqnarray*}
&\circ& \frac{\langle \phi | A | \phi \rangle }{\langle \phi | \phi \rangle}
\ge 
\frac{\langle \psi | A | \psi \rangle }{\langle \psi | \psi \rangle} . \\
&\circ&  \langle \phi | A | \phi \rangle \langle \psi | \id - A | \psi \rangle
\ge 
\langle \psi | A | \psi \rangle \langle \phi | \id - A | \phi \rangle .
\end{eqnarray*}
\end{lem}
\begin{lem}\label{le3}
we have
\begin{eqnarray*}
\left\langle  \sum_{i} \phi_i^{(n)} \right| \hat{W_{\beta}}(\rho) 
\left| \sum_{i} \phi_i^{(n)} \right\rangle 
&=& \frac{k' \cdot (k-1)!}{\pi^{(k-1)}}
\int_{\alpha}^{\frac{\pi}{ 2}} f_1( \theta_1) \cos \theta_1 \sin^{2k-3} 
\theta_1
\,d \theta_1 \\
\left\langle  \sum_{i} \phi_i^{(n)} \right| \id - \hat{W_{\beta}}(\rho) 
\left| \sum_{i} \phi_i^{(n)} \right\rangle 
&=& \frac{k' \cdot (k-1)!}{\pi^{(k-1)}}
\int_0^{\alpha} f_1( \theta_1)  \cos \theta_1 \sin^{2k-3} \theta_1\,d 
\theta_1 \\
\left \langle  \phi(0)^{(n)} \right| \hat{W_{\beta}}(\rho) 
\left| \phi(0)^{(n)} \right\rangle
&=& C \int_{\alpha}^{\frac{ \pi}{2}} 
\cos^{2n+1} \theta_1 \sin^{2k-3} \theta_1
 \,d \theta_1 \\
\left \langle  \phi(0)^{(n)} \right| \id - \hat{W_{\beta}}(\rho) 
\left| \phi(0)^{(n)} \right\rangle 
&=& C \int_0^{\alpha} 
\cos^{2n+1} \theta_1 \sin^{2k-3} \theta_1
 \,d \theta_1 .
\end{eqnarray*}
\end{lem}
\begin{pf}
$\hat{W_{\beta}}(\rho)$ is denoted as follows:
\begin{eqnarray*}
\hat{W_{\beta}}(\rho)
&= & k' \int_{\pro} W_{\beta}(\rho, \hat{\rho} ) \hat{\rho}^{(n)} 
\nu (\,d \hat{\rho}) \\
&= & k' \int_{ \{ \hat{\rho} \in \pro| \tr \hat{\rho} \rho \le \beta \} }
 \hat{\rho}^{(n)} \nu (\,d \hat{\rho}) .
\end{eqnarray*}
We obtain 
\begin{eqnarray*}
\left\langle  \sum_{i} \phi_i^{(n)} \right|  \hat{W_{\beta}}(\rho) 
\left| \sum_{i} \phi_i^{(n)} \right\rangle 
&=& 
\left\langle  \sum_{i} \phi_i^{(n)} \right|
 k' \int_{ \{ \hat{\rho} \in \pro| \tr \hat{\rho} \rho \le \beta \} }
 \hat{\rho}^{(n)} \nu (\,d \hat{\rho})  
\left| \sum_{i} \phi_i^{(n)} \right\rangle  \\
&=&
\sum_{i,j}
 k' \int_{ \{ \hat{\rho} \in \pro| \tr \hat{\rho} \rho \le \beta \} }
\left\langle \phi_i^{(n)} \right|
 \hat{\rho}^{(n)} \left| \phi_j^{(n)} \right\rangle  
 \nu (\,d \hat{\rho})  \\
&=&
\sum_{i,j}
 k' \int_{ \{ \hat{\rho} \in \pro| \tr \hat{\rho} \rho \le \beta \} }
\langle \phi_i | \hat{\rho} | \phi_j \rangle ^{n}  
 \nu (\,d \hat{\rho}) \\
&=&
\frac{k' \cdot (k-1)!}{\pi^{(k-1)}}
\int_{\alpha}^{\frac{\pi}{ 2}} f_1( \theta_1) \cos \theta_1 \sin^{2k-3} 
\theta_1
\,d \theta_1 .
\end{eqnarray*}
Similarly,
\begin{eqnarray*}
\left\langle  \sum_{i} \phi_i^{(n)} \right| \id - \hat{W_{\beta}}(\rho) 
\left| \sum_{i} \phi_i^{(n)} \right\rangle 
&=&
\sum_{i,j}
 k' \int_{ \{ \hat{\rho} \in \pro| \tr \hat{\rho} \rho \,> \beta \} }
\langle \phi_i | \hat{\rho} | \phi_j \rangle ^{n}  
 \nu (\,d \hat{\rho})  \\
&=& \frac{k' \cdot (k-1)!}{\pi^{(k-1)}} 
\int_0^{\alpha} f_1( \theta_1)  \cos \theta_1 \sin^{2k-3} \theta_1\,d 
\theta_1 \\
\left\langle \phi(0)^{(n)} \right| \hat{W_{\beta}}(\rho) 
\left| \phi(0)^{(n)} \right\rangle 
&=& 
 k' \int_{ \{ \hat{\rho} \in \pro| \tr \hat{\rho} \rho \le \beta \} }
\langle \phi(0) | \hat{\rho} | \phi(0) \rangle ^{n}  
\nu (\,d \hat{\rho})  \\
&=& C \int_{\alpha}^{\frac{ \pi}{2}} 
\cos^{2n+1} \theta_1 \sin^{2k-3} \theta_1
 \,d \theta_1 \\
\left \langle \phi(0)^{(n)} \right| \id - \hat{W_{\beta}}(\rho) 
\left| \phi(0)^{(n)} \right\rangle 
&=& 
 k' \int_{ \{ \hat{\rho} \in \pro| \tr \hat{\rho} \rho \,> \beta \} }
\langle \phi(0) | \hat{\rho} | \phi(0) \rangle ^{n}  
 \nu (\,d \hat{\rho}) \\
&=& C \int_0^{\alpha} 
\cos^{2n+1} \theta_1 \sin^{2k-3} \theta_1
 \,d \theta_1 .
\end{eqnarray*}
\end{pf}
\begin{lem}\label{lem1}
The following function $f(x)$ is monotone decreasing on $[0, \infty )$:
\begin{eqnarray*}
f(x) :=
\sum_{a=1}^m \sum_{b=1}^m
\underbrace{\int_0^{2 \pi} \ldots \int_0^{2 \pi} }_{k}
\Bigl( c_a^0 e^{i d_a^0}+ x \sum_{j=1}^{k} e^{i (\theta_j + d_a^j)} c_a^j 
\Bigr)^n
\Bigl( c_b^0 e^{i d_b^0}+ x \sum_{j=1}^{k} e^{-i (\theta_j + d_b^j)} c_b^j 
\Bigr)^n
\,d \theta_1 \cdots \,d \theta_k .
\end{eqnarray*}
where $c_n^j,d_n^j$ are any real numbers.
\end{lem}
\begin{pf}
The set $K^m_n$ is defined as follows:
\begin{eqnarray*}
K^m_n := \left\{ I=(I_1, \cdots , I_m ) \in ({\bf N}^{+,0})^m \left|
 \sum_{j=1}^m I_j = n \right.\right\}.
\end{eqnarray*}
The number $C(I)$ is defined for $I \in K^m_n$ as sufficing the following 
condition:
\begin{eqnarray*}
\left(\sum_{j=1}^m x_j \right)^n 
= \sum_{I \in K^m_n} C(I) x_1^{I_0} \ldots x_m^{I_m}.
\end{eqnarray*}
Therefore, 
\begin{eqnarray*}
\left( c_a^0 + x \sum_{j=1}^{k} e^{i (\theta_j + d_a^j)} c_a^j \right)^n
=
\sum_{I \in K^{k+1}_n } C(I) e^{i d_a^0}(c_a^0)^{I_0} 
e^{i I_1(\theta_1 + d_a^1)} (c_a^1)^{I_1} \ldots
 e^{i I_d(\theta_k + d_a^k)} (c_a^k)^{I_k} x^{n-I_0}.
\end{eqnarray*}
Thus,
\begin{eqnarray*}
&~&f(x) \\
&=&
\sum_{a=1}^m \sum_{b=1}^m
( 2 \pi)^k
\sum_{I} C(I) e^{i I_0(d_a^0 - d_b^0)} (c_a^0 c_b^0)^{I_0} 
e^{i I_1(d_a^1 - d_b^1)} (c_a^1 c_b^1)^{I_1} \ldots
 e^{i I_k(d_a^k - d_b^k)} (c_a^k c_b^k)^{I_k} x^{2n-2I_0} \\
&=&
( 2 \pi)^k
\sum_{I} C(I) 
\sum_{a=1}^m \sum_{b=1}^m
e^{i (\sum_{j=0}^k I_i d_a^i - \sum_{j=0}^k I_i d_b^i )}
 (c_a^0 )^{I_0} \ldots (c_a^k )^{I_k}
(c_b^0 )^{I_0} \ldots (c_b^k )^{I_k} x^{2n-2I_0} \\
&=&
( 2 \pi)^k
\sum_{I} C(I) D(I)
x^{2n-2I_0},
\end{eqnarray*}
where
\begin{eqnarray*}
D(I) :=
\sum_{a=1}^m \sum_{b=1}^m
e^{i (\sum_{j=0}^k I_i d_a^i - \sum_{j=0}^k I_i d_b^i )}
 (c_a^0 )^{I_0} \ldots (c_a^k )^{I_k}
(c_b^0 )^{I_0} \ldots (c_b^k )^{I_k} .
\end{eqnarray*}
It is sufficient to show $D(I) \ge 0$.
Letting
\begin{eqnarray*}
v_a &:=& (c_a^0 )^{I_0} \ldots (c_a^k )^{I_k} \\
y_a &:=& \sum_{j=0}^k I_i d_a^i \\
w_{a,b} &:=& \cos (y_a - y_b),
\end{eqnarray*}
we have
\begin{eqnarray*}
D(I)=\sum_{a=1}^m \sum_{b=1}^m
v_a w_{a,b} v_b.
\end{eqnarray*}
Then
\begin{eqnarray*}
w_{a,b} = \cos (y_a - y_b)
= \cos y_a \cos y_b + \sin y_a \sin y_b .
\end{eqnarray*}
As $\{ \cos y_a \cos y_b \}$ and $\{ \sin y_a \sin y_b \}$ 
are nonnegative,
$\{ w_{a,b} \}$ is nonnegative matrix.
Therefore,
we obtain $D(I) \ge 0$.
\end{pf}
\section{Proof of Lemma \protect{\ref{bunpu}}}\label{pf1}
\begin{eqnarray*}
{\cal D}^{W,(n)}(\Pi_n) &=&
\int_{\pro} h( d_{fs}(\rho,\hat{\rho}) \tr 
(\Pi_n (\,d \hat{\rho}) \rho^{(n)}) \\
&=&
\int_{\pro} h( \theta_1 ) 
{n+k-1 \choose k-1} | \langle \phi(\theta)^{(n)} 
| \phi (0)^{(n)} \rangle |^2 \nu (\,d \theta) \\
&=&
\int_0^{\frac{\pi}{2}}  h( \theta_1 ) 
{n+k-1 \choose k-1}\frac{ (k-1)! }{\pi^{k-1}} \cos^{2n+1} \theta_1  
\sin^{2k-3} \theta_1 \,d \theta_1 \\
&~& \times 
\underbrace{\int_0^{2 \pi} \cdots \int_0^{2 \pi}}_{k-1}  
\underbrace{\int_0^{\frac{\pi}{2}} \cdots \int_0^{\frac{\pi}{2}}}_{k-2}
\sin^{2k-5} \theta_2 \cdots \sin \theta_{k-1} 
\cos \theta_2 \cdots \cos \theta_{k-1} 
\,d \theta_2 \cdots \,d \theta_{2k-2} \\
&=&
\int_0^{\frac{\pi}{2}} h( \theta_1 ) 
{n+k-1 \choose k-1}\frac{ (k-1)! }{\pi^{k-1}} \cos^{2n+1} \theta_1  
\sin^{2k-3} \theta_1 \,d \theta_1 \\
&~& \times 
\underbrace{\int_0^1 x^{2k-5} \,d x \cdots \int_0^1 x \,d x }_{k-2}
\cdot (2 \pi)^{k-1} \\
&=&
\int_0^{\frac{\pi}{2}} h( \theta ) 
{n+k-1 \choose k-1}\frac{ (k-1)! }{\pi^{k-1}} \cos^{2n+1} \theta 
\sin^{2k-3} \theta \,d \theta
\frac{(2 \pi)^{k-1}}{2^{k-2}(k-2)!} \\
&=& 2(k-1) {n+k-1 \choose k-1}
\int_0^{\frac{\pi}{2}} h( \theta ) 
\cos^{2n+1} \theta 
\sin^{2k-3} \theta \,d \theta .
\end{eqnarray*}
The proof is complete.
\section{Proof of Thereon \protect\ref{th61}}\label{pf2}
Definition \ref{fsb} and Lemma \ref{bunpu} means that:
 \begin{eqnarray}
 {\cal D}^{d_b^{\gamma},(n)}(\Pi_n)
 =2(k-1){n+k-1 \choose k-1} \int_0^{\frac{\pi}{2}} \cos^{2n+1} \theta 
 \sin^{2k-3+ \gamma} \theta \,d \theta . \label{kan}
 \end{eqnarray}
Since 
 \begin{eqnarray*}
 \int_0^{\frac{\pi}{2}} \cos^{x} \theta 
 \sin^{y} \theta \,d \theta
 = \frac{\Gamma (\frac{x+1}{2}) \Gamma( \frac{y+1}{2})}
 {2 \Gamma (\frac{x+y}{2}+1)} \quad\forall x,y \in {\bf R},
 \end{eqnarray*}
we have
 \begin{eqnarray}
 {\cal D}^{d_b^{\gamma},(n)}(\Pi_n) 
 &=& 
 2 (k-1) {n+k-1 \choose k-1}
 \frac{\Gamma (n+1 ) \Gamma( k-1+\gamma/2)}
 {  \Gamma (n + k +\gamma/2)} \nonumber  \\
 &=& 
 \frac{\Gamma (n+1 ) \Gamma( k-1+\gamma/2)\Gamma (n + k )}
 {  \Gamma (n + k +\gamma/2)\Gamma (n + 1) \Gamma( k-1)} \nonumber  \\
 &=& \frac{\Gamma (n + k )}{ \Gamma (n + k +\gamma/2)}
 \frac{\Gamma( k-1+\gamma/2)}{\Gamma( k-1)} . \label{keu}
\end{eqnarray}
Therefore we obtain (\ref{kere}) from the following
formula of $\Gamma$ function:
 \begin{eqnarray*}
 \lim_{n \to \infty} \frac{\Gamma (n +x) }{\Gamma(n)n^x} = 1.
 \end{eqnarray*}
 Letting $\gamma:=2$, we obtain 
 \begin{eqnarray*}
 {\cal D}^{d_b^{2},(n)}(\Pi_n) 
 =\frac{\Gamma (n + k )}{ \Gamma (n + k +1)}
 \frac{\Gamma( k-1+1)}{\Gamma( k-1)} 
 = \frac{k-1}{ n + k } .
 \end{eqnarray*}
Thus, we get (\ref{keha}).

Next, we will prove (\ref{bun1}).
$\theta^2$ can be expanded as:
\begin{eqnarray*}
\theta^2 = \sum_{i=0}^{\infty} \frac{(2i-2)!!}{(2i-1)!!}
\frac{\sin^{2i} \theta}{i},
\end{eqnarray*}
where we put $(2n)!!=2n(2n-2) \cdots 4 \cdot 2,(2n-1)!!=(2n-1)(2n-3) 
\cdots 3 \cdot 1
,0!!=(-1)!!=1$.
From (\ref{keu}) we have
\begin{eqnarray*}
 {\cal D}^{d_{fs}^2,(n)}(\Pi_n) 
&=&  \sum_{i=0}^{\infty} \frac{(2i-2)!!}{(2i-1)!!i}
{\cal D}^{d_{b}^{2i},(n)}(\Pi_n) \\
&=&  \sum_{i=0}^{\infty} \frac{(2i-2)!!}{(2i-1)!!i}
\prod_{j=0}^{i-1} \frac{k-1+j}{n+k+j} \\
&\cong& (k-1) \frac{1}{n} - \frac{2}{3}k(k-1)\frac{1}{n^2}
+ k(k-1)\frac{23k-7}{45} \frac{1}{n^3}.
\end{eqnarray*}
Thus we obtain (\ref{bun1}).
Lemma \ref{bunpu} derives that 
\begin{eqnarray}
 &~& \log 
 \Pr_{\Pi_n}^{\rho^{(n)}} \{ \hat{\rho} \in \pro 
| d_{fs}(\rho, \hat{\rho}) \ge \epsilon \} \nonumber\\
&=& \log 
\left( (k-1){n+k-1 \choose k-1} 
\int_0 ^{\cos^2 \epsilon} x^n ( 1- x)^{k-2} \,d x \right) .
\end{eqnarray}
Therefore, it is sufficient for (\ref{zenkin5}) to show that
\begin{eqnarray}
\log {n+k-1 \choose k-1} 
&\cong& (k-1)\log n - \log (k-1) ! + \frac{1}{n} \frac{(k-1)k}{2} 
\label{choo} \\
\log \left( \int_0 ^{\cos^2 \epsilon} x^n ( 1- x)^{k-2} \,d x \right)
&\cong& 2n \log \cos \epsilon -\log n
+ 2 (k-2) \log \sin \epsilon -2 \log \cos \epsilon \nonumber \\
&~&- \frac{1}{n} ( 1+ (k-2) \cot^2 \epsilon ) .\label{int}
\end{eqnarray}
The left hand of (\ref{choo}) is calculated as:
\begin{eqnarray*}
\log {n+k-1 \choose k-1}
&=& \sum_{i=0}^{k-1} \log \frac{n+i}{i} \\
&=&  (k-1)\log n - \log (k-1) !
+ \sum_{i=1}^{k-1} \log ( 1+ \frac{i}{n}) \\
&\cong & (k-1)\log n - \log (k-1) !
+ \sum_{i=1}^{k-1}  \frac{i}{n}\\
&=&  (k-1)\log n - \log (k-1) !
+\frac{1}{n} \frac{(k-1)k}{2} .
\end{eqnarray*}
Therefore, we have (\ref{choo}).
The left hand of (\ref{int}) is calculated as:
\begin{eqnarray*}
&~&\log \left( \int_0 ^{\cos^2 \epsilon} x^n ( 1- x)^{k-2} \,d x
\right)
- 2n \log \cos \epsilon  \\
&=& \log \left( \int_0 ^{\cos^2 \epsilon} \left( \frac{x}{\cos^2 \epsilon} 
\right)^n
( 1- x)^{k-2} \,d x \right) \\
&=& \log \left( \int_0^1 x^n \left ( 1- \cos^2 x \right ) ^{k-2} 
\frac{1}{\cos^2 \epsilon}
\,d x  \right) \\
&=& -2 \log \cos \epsilon + \log \left( \sum_{i=0}^{k-2} {k-2 \choose i} 
\left ( - \cos^2 \epsilon \right )^i \frac{1}{n+i+1} \right) \\
&=& -2 \log \cos \epsilon  -\log n +\log \left( \sum_{i=0}^{k-2} 
{k-2 \choose i} 
\left ( - \cos^2 \epsilon \right )^i \frac{1}{1+\frac{i+1}{n}} \right) \\
&\cong& -2 \log \cos \epsilon  -\log n +\log \left( \sum_{i=0}^{k-2} 
{k-2 \choose i} 
\left ( - \cos^2 \epsilon \right )^i \left(1-\frac{i+1}{n} \right) \right) \\
&=& -2 \log \cos \epsilon  -\log n +
\log \left( \left (1- \cos^2 \epsilon \right )^{k-2} 
- \frac{1}{n} \left (1- \cos^2 \epsilon \right )^{k-3}
\left ( 1- ( k-1)\cos^2 \epsilon \right ) \right) \\
&=& -2 \log \cos \epsilon  -\log n +
\log \left (1- \cos^2 \epsilon \right )^{k-2}  
+\log \left( 1 - \frac{1}{n} 
\frac{1- ( k-1)\cos^2 \epsilon }{1- \cos^2 \epsilon}
 \right) \\
&\cong& -2 \log \cos \epsilon  -\log n +
(k-2)\log \left (1- \cos^2 \epsilon \right )
- \frac{1}{n} 
\frac{1- ( k-1)\cos^2 \epsilon }{1- \cos^2 \epsilon}.
\end{eqnarray*}
We obtain (\ref{int}).
\section{Proof of Lemma \protect\ref{kaka}}\label{pf3}
From the symmetry of $\pro$ and $\Pi_m$,
we may assume that $\rho_1 = | \phi_0 \rangle \langle \phi_0 |,
\rho_2 = | \phi_{\epsilon} \rangle \langle \phi_{\epsilon} |$.
First, we consider the case of $k=2$.
For the following calculation, we prepare the following equations:
\begin{eqnarray}
&~& |\langle \phi_\epsilon | \phi(\theta) \rangle |^{2n} \nonumber \\
&~ &= (\cos^2 \epsilon \cos^2 \theta_1 + \sin^2 \theta_1 \sin^2 \epsilon+
 2 \cos \epsilon \sin \epsilon \cos \theta_1 \sin \theta_1 \cos \theta_2 )^n
 \label{keti1} \\
&~& \int_0^{2 \pi} \log ( 1 + 2 a \cos \theta + a^2) \,d \theta
= 4 \pi \psi (| a|) \log | a |  ,\label{keti2}
\end{eqnarray}
where the function $\psi$ is defined as: 
\begin{eqnarray*}
\psi(x)=
\left\{
\begin{array}{@{\,}ll}
1 & x \ge 1 \\
0 & x \,< 0
\end{array}
\right. .
\end{eqnarray*}
Paying attention to (\ref{alpha}) and (\ref{beta}),
we have 
\begin{eqnarray}
&~& \frac{-D_{\Pi_m}(\rho_1^{(m)} \| \rho_2^{(m)}) - m \log \cos^2 \epsilon}
{m} \nonumber \\
&=& -\frac{1}{m}
\left( (m+1)
\int_{\pro}
\log \frac{ | \langle \phi_0 | \phi(\theta) \rangle|^{2m}}
{ | \langle \phi_\epsilon | \phi(\theta) \rangle|^{2m}}
| \langle \phi_0 | \phi(\theta) \rangle|^{2m} \nu (\,d \theta)
+ m \log \cos \epsilon \right) \nonumber \\
&=& -\frac{2(m+1)}{\pi} \ \nonumber \\\
&~& \cdot \int_0^{2 \pi} \int_0^{\frac{\pi}{2}}
\log \left( \frac{ \cos^2 \theta_1 \cos^2 \epsilon}
{ (\cos \theta_1 \cos \epsilon 
+ \sin \theta_1 \cos \theta_2 \sin \epsilon)^2 
+ ( \sin \theta_1 \sin \theta_2 \sin \epsilon)^2 } \right) \nonumber \\
&~& \cdot \cos^{2m+1} \theta_1 \sin \theta_1 
\,d \theta_1 \,d \theta_2 \nonumber \\
&=& \frac{(m+1)}{\pi}  \nonumber \\
&~& \cdot \int_0^{\frac{\pi}{2}}
\int_0^{2 \pi} \log  \left(1 + 2 \tan \theta_1 \tan \epsilon \cos \theta_2
+ (\tan \theta_1 \tan \epsilon )^2 \right)\,d \theta_2
\cos^{2m+1} \theta_1 \sin \theta_1 
\,d \theta_1  \nonumber \\
&=& \frac{(m+1)}{\pi} 
\int_0^{\frac{\pi}{2}}
4 \pi \log (\tan \theta_1 \tan \epsilon )
\psi (\tan \theta_1 \tan \epsilon )
\cos^{2m+1} \theta_1 \sin \theta_1 
\,d \theta_1 \nonumber \\
&=& \frac{(m+1)}{\pi} 
\int_{\frac{\pi}{2}-\epsilon}^{\frac{\pi}{2}}
4 \pi \log (\tan \theta_1 \tan \epsilon )
\cos^{2m+1} \theta_1 \sin \theta_1 
\,d \theta_1 \nonumber \\
&=&  2(m+1) 
\int_{\frac{\pi}{2}-\epsilon}^{\frac{\pi}{2}}
\log \left (\tan^2 \theta_1 \tan^2 \epsilon \right )
\cos^{2m+1} \theta_1 \sin \theta_1 
\,d \theta_1   \nonumber \\
&=& (m+1) 
\int_0^{\sin^2 \epsilon}
\log \left(\frac{1-x}{x} \tan^2 \epsilon \right)
x^m \,d x \label{kaki1}.
\end{eqnarray}
Substitute $a = \tan^2 \epsilon$ in (\ref{he}) of Lemma \ref{hohe} 
, then 
\begin{eqnarray}
(m+1) 
\int_0^{\sin^2 \epsilon}
x^m \log \left(\frac{1-x}{x} \tan^2 \epsilon \right)
\,d x  
= -\log \cos^2 \epsilon - \sum_{i=1}^m \frac{\sin^{2i} \epsilon}{i} 
 . \label{kaki2}
\end{eqnarray}
From (\ref{kaki1}) and (\ref{kaki2}), we have 
\begin{eqnarray}
 \frac{D_{\Pi_m}(\rho_1^{(m)} \| \rho_2^{(m)}) }{m}
= \sum_{i=1}^m \frac{\sin^{2i} \epsilon }{i} . \label{kaki3}
\end{eqnarray}
Therefore, we can prove (\ref{kiuu}).

Next, we consider the case of $k \ge 3$.
In this case, we have:
\begin{eqnarray}
&~& |\langle \phi_\epsilon | \phi(\theta) \rangle |^{2n} \nonumber \\
&=& (\cos^2 \epsilon \cos^2 \theta_1 + \sin^2 \theta_1 \cos^2 \theta_2 
\sin^2 \epsilon+
 2 \cos \epsilon \sin \epsilon \cos \theta_1 \sin \theta_1 \cos \theta_2 
\cos \theta_k)^n .\label{keti} 
\end{eqnarray}
Paying attention to (\ref{alpha}),(\ref{keti2}) and Lemma \ref{ha},
we can calculate as:
\begin{eqnarray}
&~&  \frac{-D_{\Pi_m}(\rho_1^{(m)} \| \rho_2^{(m)}) -m \log \cos^2 \epsilon}
{m} \nonumber \\
&=& -\frac{1}{m} \left( {m+k-1 \choose k-1}
\int_{\pro}
\log \left( \frac{ | \langle \phi_0 | \phi(\theta) \rangle|^{2m}}
{ | \langle \phi_\epsilon | \phi(\theta) \rangle|^{2m}} \right)
| \langle \phi_0 | \phi(\theta) \rangle|^{2m} \nu (\,d \theta)
+ 2m \log \cos \epsilon \right) \nonumber \\
&=& -\frac{2(k-1)(k-2)}{\pi} {m+k-1 \choose k-1} \nonumber \\
&~& \cdot \int_0^{2 \pi} \int_0^{\frac{\pi}{2}}\int_0^{\frac{\pi}{2}}
\log \left( \frac{ \cos^2 \theta_1 \cos^2 \epsilon}
{ (\cos^2 \epsilon \cos^2 \theta_1 + \sin^2 \theta_1 \cos^2 \theta_2 
\sin^2 \epsilon+
 2 \cos \epsilon \sin \epsilon \cos \theta_1 \sin \theta_1 \cos \theta_2 
\cos \theta_k)
 }\right)\nonumber \\
&~& \cdot \cos^{2m+1} \theta_1 \sin^{2k-3} \theta_1 \cos \theta_2
 \sin^{2k-5} \theta_2
\,d \theta_1 \,d \theta_2 \,d \theta_k \nonumber \\
&=&  \frac{2(k-1)(k-2)}{\pi} {m+k-1 \choose k-1} 
\int_0^{\frac{\pi}{2}}\int_0^{\frac{\pi}{2}} \nonumber \\
&~& \cdot \int_0^{2 \pi} \log  \left(1 + 2 \tan \theta_1 \cos \theta_2 
\tan \epsilon \cos \theta_k
+ (\tan \theta_1 \cos \theta_2 \tan \epsilon )^2 \right)\,d \theta_k 
\nonumber \\
&~& \cdot \cos^{2m+1} \theta_1 \sin^{2k-3} \theta_1 \cos \theta_2 
\sin^{2k-5} \theta_2
\,d \theta_1 \,d \theta_2 \nonumber \\
&=& \frac{2(k-1)(k-2)}{\pi} {m+k-1 \choose k-1} 
\int_0^{\frac{\pi}{2}}\int_0^{\frac{\pi}{2}}
4 \pi \psi (\tan \theta_1 \cos \theta_2 \tan \epsilon ) \nonumber \\
&~& \cdot \log \left( \tan \theta_1 \cos \theta_2 \tan \epsilon \right)  
\cos^{2m+1} \theta_1 \sin^{2k-3} \theta_1 \cos \theta_2 \sin^{2k-5} \theta_2
\,d \theta_1 \,d \theta_2 . \label{ni}
\end{eqnarray}
Substitute that $a:=\tan^2 \epsilon, s:=\sin^2 \theta_2, y:=\cos^2 \theta_1$,
then the condition \par
$ \tan \theta_1 \cos \theta_2 \tan \epsilon \ge 1$
turns into the following conditions:
\begin{eqnarray*}
1- \frac{y}{a(1-y)} \ge x \ge 0 ,~ \frac{a}{1+a} \ge y \ge 0
\end{eqnarray*}
Using (\ref{ni}) and (\ref{ho}), we have
\begin{align}
&\quad  \frac{-D_{\Pi_m}(\rho_1^{(m)} \| \rho_2^{(m)}) 
-m \log \cos^2 \epsilon}{m} 
\nonumber \\
&= (k-1)(k-2) { m+k-1 \choose k-1} 
\int_0^{\frac{a}{1+a}} \left(
\int_0^{1- \frac{y}{a(1-y)} }
x^{k-3} \log\left( (1-x) \frac{a(1-y)}{y} \right)
\,d x \right)
y^m (1-y)^{k-2} \,d y \nonumber \\
&= (k-1){ m+k-1 \choose k-1} 
\int_0^{\frac{a}{1+a}} \left(
-\log \left( \frac{y}{a(1-y)} \right)
- \sum_{i=1}^{k-2} \frac{1}{i}\left( \frac{a-(1+a)y}{a(1-y)} \right)^i
\right)
y^m (1-y)^{k-2} \,d y \nonumber \\
&= -(k-1){ m+k-1 \choose k-1}
\int_0^{\frac{a}{1+a}} 
\log \left( \frac{y}{a(1-y)} \right)y^m (1-y)^{k-2} \,d y  
- (k-1){ m+k-1 \choose k-1} f\left( \frac{a}{1+a}\right ) ,\label{ti}
\end{align}
where $f(x)$ is defined as:
\begin{eqnarray*}
f(x):= 
\int_0^{x} \sum_{i=1}^{k-2}\frac{1}{i}\left( 1-\frac{y}{x} \right)^i
y^m (1-y)^{k-2-i} \,d y .
\end{eqnarray*}
From Lemma \ref{to}, the derivative of $f(x)$ can be calculated as:
\begin{eqnarray}
f'(x)
&=&
\int_0^{x} \frac{y}{x^2}
\left( \sum_{i=1}^{k-2}\left( \frac{1-\frac{y}{x}}{1-y} \right)^{i-1}
\right)
y^m (1-y)^{k-3} \,d y \nonumber \\
&=&
\int_0^{x} \frac{y}{x^2}
\left(
\frac{1- \left( \frac{1-\frac{y}{x}}{1-y} \right)^{k-2}}
{1- \frac{1-\frac{y}{x}}{1-y} }
\right)
y^m (1-y)^{k-3} \,d y \nonumber \\
&=&
\frac{1}{x(1-x)}
\int_0^{x} 
\left(
1- \left( \frac{1-\frac{y}{x}}{1-y} \right)^{k-2}
\right)
y^m (1-y)^{k-2} \,d y \nonumber \\
&=&
\frac{1}{x(1-x)}
\int_0^{x} 
y^m (1-y)^{k-2} \,d y 
-\frac{1}{x(1-x)}
\int_0^{x} 
\left( 1-\frac{y}{x} \right)^{k-2}
y^m \,d y \nonumber \\
&=&
\frac{1}{x(1-x)}
\sum_{i=0}^{k-2} {k-2 \choose i} 
\int_0^{x} (-1)^i y^{m+i} \,d y
-\frac{x^{m+1}}{x(1-x)}
\int_0^{1} 
\left( 1-t \right)^{k-2}
t^m \,d t \nonumber \\
&=&
\frac{x^m}{x(1-x)}
\sum_{i=0}^{k-2} {k-2 \choose i} 
\frac{(-x)^i}{m+i+1}
-\frac{x^{m+1}}{x(1-x)}
{m+k-1 \choose k-2}^{-1} \frac{1}{m+1}. \label{ri}
\end{eqnarray}
By (\ref{he}) and Lemma \ref{to},
the first term of (\ref{ti}) is calculated as:
\begin{eqnarray}
&~& -(k-1){ m+k-1 \choose k-1} 
\int_0^{\frac{a}{1+a}} 
y^m (1-y)^{k-2} \log \left( \frac{y}{a(1-y)} \right)
\,d y \nonumber \\
&=& -(k-1){ m+k-1 \choose k-1} 
\sum_{i=0}^{k-2} { k-2 \choose i} (-1)^i
\int_0^{\frac{a}{1+a}} 
y^{m+i} \log \left( \frac{y}{a(1-y)} \right)
\,d y \nonumber \\
&=& -(k-1){ m+k-1 \choose k-1} 
\sum_{i=0}^{k-2} { k-2 \choose i} \frac{(-1)^i}{m+i+1} 
\left(
- \log (1+a) + 
\sum_{j=1}^{m+i} \frac{1}{j} \left( \frac{a}{1+a} \right)^j 
\right) \nonumber \\
&=& -(k-1){ m+k-1 \choose k-1} 
\sum_{i=0}^{k-2} { k-2 \choose i} \frac{(-1)^i}{m+i+1} \nonumber \\
&~& \times
\left(
- \log (1+a) + 
\sum_{j=1}^{m} \frac{1}{j} \left( \frac{a}{1+a} \right)^j +
\sum_{j=m+1}^{m+i} \frac{1}{j} 
\left( \frac{a}{1+a} \right)^j \right) \nonumber \\
&=& -(k-1){ m+k-1 \choose k-1} 
\left( \sum_{i=0}^{k-2} { k-2 \choose i} \frac{(-1)^i}{m+i+1} \right) 
 \left(
- \log (1+a) + 
\sum_{j=1}^{m} \frac{1}{j} \left( \frac{a}{1+a} \right)^j \right)\nonumber \\
&~& -(k-1){ m+k-1 \choose k-1} 
\sum_{i=0}^{k-2} \sum_{j=1}^{i} { k-2 \choose i} \frac{(-1)^i }{(m+i+1)(j+m)}
\left( \frac{a}{1+a} \right)^{j+m} \nonumber \\
&=& -(k-1){ m+k-1 \choose k-1} 
\left( { m+k-1 \choose k-2 }^{-1} \frac{1}{m+1}\right)
\left(
- \log (1+a) + 
\sum_{j=1}^{m} \frac{1}{j} \left( \frac{a}{1+a} \right)^j \right) \nonumber \\
&~& +(k-1){ m+k-1 \choose k-1} g\left(\frac{a}{1+a} \right)\nonumber  \\
&=&
\log (1+a) -
\sum_{j=1}^{m} \frac{1}{j} \left( \frac{a}{1+a} \right)^j 
-(k-1){ m+k-1 \choose k-1} g\left(\frac{a}{1+a} \right) , \label{nu}
\end{eqnarray}
where $g(x)$ is defined as:
\begin{eqnarray*}
g(x):=
\sum_{i=0}^{k-2} \sum_{j=1}^{i} { k-2 \choose i} \frac{(-1)^i }{(m+i+1)(j+m)}
x^{j+m} .
\end{eqnarray*}
By Lemma \ref{to}, the derivative of $g(x)$ is calculated as:
\begin{eqnarray}
g'(x)
&=&\sum_{i=0}^{k-2} { k-2 \choose i} \frac{(-1)^i }{(m+i+1)}
\sum_{j=1}^{i}x^{j+m-1} \nonumber \\
&=&\sum_{i=0}^{k-2} { k-2 \choose i} \frac{(-1)^i }{(m+i+1)}
x^m \frac{1-x^i}{1-x}\nonumber \\
&=& \frac{x^m}{1-x} \sum_{i=0}^{k-2} { k-2 \choose i} \frac{(-1)^i }{(m+i+1)}
- \frac{x^m}{1-x} \sum_{i=0}^{k-2} 
{ k-2 \choose i} \frac{(-x)^i }{(m+i+1)} \nonumber \\
&=& \frac{x^m}{1-x} {m+k-1 \choose k-2}^{-1} \frac{1}{m+1}
- \frac{x^m}{1-x} \sum_{i=0}^{k-2} { k-2 \choose i} \frac{(-x)^i }{(m+i+1)}. 
\label{ru}
\end{eqnarray} 
From (\ref{ri}) and (\ref{ru}), we have $f'(x)=-g'(x)$.
The definitions of $f(x)$ and $g(x)$ means
that $f(0)=g(0)=0$.
Then we obtain $f(x)=-g(x)$.
By (\ref{ti}) and (\ref{nu}), we have
\begin{eqnarray*}
&~&  \frac{-D_{\Pi_m}(\rho_1^{(m)} \| \rho_2^{(m)}) 
-2m \log \cos \epsilon}{m} \\
&=& (k-1)(k-2) { m+k-1 \choose k-1} 
\int_0^{\frac{a}{1+a}} \left(
\int_0^{1- \frac{y}{a(1-y)}}
x^{k-3} \log\left( (1-x) \frac{a(1-y)}{y} \right)
\,d x \right)
y^m (1-y)^{k-2} \,d y \\
&=&  
\log (1+a) - 
\sum_{j=1}^{m} \frac{1}{j} \left( \frac{a}{1+a} \right)^j \\
&~& -(k-1){ m+k-1 \choose k-1} 
\left( g\left(\frac{a}{1+a} \right)
+ f\left(\frac{a}{1+a} \right) \right)\\
&=& 
\log (1+a)  -
\sum_{j=1}^{m} \frac{1}{j} \left( \frac{a}{1+a} \right)^j \\
&=& -\log \cos^2 \epsilon
-\sum_{j=1}^{m} \frac{1}{j} \sin^{2j} \epsilon .
\end{eqnarray*}
Then, we obtain:
\begin{eqnarray*}
\frac{D_{\Pi_m}(\rho_1^{(m)} \| \rho_2^{(m)})}{m} 
= \sum_{j=1}^{m} \frac{\sin^{2i} \epsilon}{i} .
\end{eqnarray*}
We proved (\ref{kiuu}).\par
Next we will prove (\ref{kaka1}).
 We consider the tangent space $T_{\rho}\pro$ at 
$\rho:= | \phi (0) \rangle \langle \phi(0)|$.
 If $c(t)$ is a curve on $\pro$ such that $c(0)= \rho$,
 $\dot{c}$ denotes the element of $T_{\rho}\pro$ defined by $c(t)$.
 the Fubini-Study metric $g_{fs}$ is defined as:
 \begin{eqnarray*}
 g_{fs}(\dot{c},\dot{c}) 
:= \left( \lim_{t \to 0} \frac{d_{fs}(c(0),c(t))}{t} \right)^2
\end{eqnarray*}
Therefore, it is sufficient to show that
\begin{eqnarray*}
J_{\Pi_n}^{\rho}
= 2 n g_{fs}.
\end{eqnarray*}
Let $c(t):= | \phi_t \rangle \langle \phi_t |,
\phi_t:= \phi( t, 0,\ldots ,0)$. (See the equation (\ref{tei5}).)
Because $g_{fs}(\dot{c},\dot{c})=1$, 
it is sufficient to prove that
\begin{eqnarray*}
J_{\Pi_n}^{\rho}(\dot{c},\dot{c})= 2n.
\end{eqnarray*}
We assume that $k \ge 3$.
From (\ref{keti}), we have
\begin{eqnarray}
 \left( \left. \frac{\,d }{\,d t}\log 
\left( |\langle \phi_t | \phi(\theta) \rangle |^{2n} \right) 
\right |_{t=0} \right)^2
 |\langle \phi_0 | \phi(\theta) \rangle |^{2n} 
 = 4 n^2 \cos^{2n-2} \theta_1 \sin^2 \theta_1 \cos^2 \theta_2 \cos^2 \theta_k .
\label{wo}
\end{eqnarray}
By (\ref{wo}) and (\ref{ha1}), we have:
\begin{eqnarray}
 &~& {m+k-1 \choose k-1} 
 \int_{\pro}
\left (\frac{\,d }{\,d t}\log \left. 
\left( |\langle \phi_t | \phi(\theta) \rangle |^{2m}\right)\right 
|_{t=0} \right)^2
 |\langle \phi_0 | \phi(\theta) \rangle |^{2m}
\nu(\,d \theta )  \nonumber \\
 &=&  \frac{2(k-1)(k-2)}{\pi}{m+k-1 \choose k-1} 4 m^2 
 \int_{0}^{\frac{ \pi}{2}} \cos^{2n-1} \theta_1 \sin^{2k-1} \theta_1 
\,d \theta_1 \nonumber \\
&~& \cdot \int_{0}^{\frac{ \pi}{2}} \cos^3 \theta_2
\sin^{2k-5} \theta_2 \,d \theta_2
 \int_{0}^{2 \pi} \cos^2 \theta_k \,d \theta_k \nonumber \\
 &=&  \frac{2(k-1)(k-2)}{\pi}{m+k-1 \choose k-1} 4 m^2 
 \frac{ (m-1) ! (k-1)!}{2(m+k-1)!}
 \frac{ 1! (k-3)!}{2 (k-1)!}
 \pi \nonumber \\
 &=& 2m. \label{keke}
\end{eqnarray}
We get (\ref{kaka1}).
In the case of $k=2$, similarly we can prove (\ref{kaka1}).
\begin{lem}\label{ha}
If $k \ge 3$, then we have
\begin{eqnarray}
&~&\int_{\pro} f(\theta_1, \theta_2, \theta_k ) \nu (\,d \theta) \nonumber \\
&=& \frac{2 (k-1)(k-2)}{\pi}
\int_0^{2 \pi} \int_0^{\frac{\pi}{2}}\int_0^{\frac{\pi}{2}}
f(\theta_1, \theta_2, \theta_k ) 
\cos \theta_1 \sin^{2k-3} \theta_1 \,d \theta_1 
\cos \theta_2 \sin^{2k-5} \theta_2 \,d \theta_2 \,d \theta_k .\label{ha1}
\end{eqnarray}
\end{lem}
\begin{pf}
From (\ref{beta}) the left hand of (\ref{ha1}) is calculated as:
\begin{eqnarray*}
&~&
\int_{\pro} f(\theta_1, \theta_2, \theta_k ) \nu (\,d \theta) \\
&=& \frac{(k-1)!}{\pi^{k-1}}
\int_0^{2 \pi} \int_0^{\frac{\pi}{2}}\int_0^{\frac{\pi}{2}}
f(\theta_1, \theta_2, \theta_k ) 
\cos \theta_1 \sin^{2k-3} \theta_1 \,d \theta_1 
\cos \theta_2 \sin^{2k-5} \theta_2 \,d \theta_2 \,d \theta_k \\
&~& \times 
\underbrace{\int_0^{\frac{\pi}{2}} \cdots \int_0^{\frac{\pi}{2}}}_{k-3}
\sin^{2k-7} \theta_3 \cdots \sin \theta_{k-1} 
\cos \theta_3 \cdots \cos \theta_{k-1} 
\,d \theta_2 \cdots \,d \theta_{k-1} 
\underbrace{\int_0^{2 \pi} \cdots \int_0^{2 \pi}}_{k-2}  
\,d \theta_{k+1} \cdots \,d \theta_{2k-2} \\
&=& 
\int_0^{2 \pi} \int_0^{\frac{\pi}{2}}\int_0^{\frac{\pi}{2}}
f(\theta_1, \theta_2, \theta_k ) 
\cos \theta_1 \sin^{2k-3} \theta_1 \,d \theta_1 
\cos \theta_2 \sin^{2k-5} \theta_2 \,d \theta_2 \,d \theta_k \\
&~& \times \frac{(k-1)!}{\pi^{k-1}}
\underbrace{\int_0^1 x^{2k-7} \,d x \cdots \int_0^1 x \,d x }_{k-3}
\cdot (2 \pi)^{k-2} \\
&=& 
\int_0^{2 \pi} \int_0^{\frac{\pi}{2}}\int_0^{\frac{\pi}{2}}
f(\theta_1, \theta_2, \theta_k ) 
\cos \theta_1 \sin^{2k-3} \theta_1 \,d \theta_1 
\cos \theta_2 \sin^{2k-5} \theta_2 \,d \theta_2 \,d \theta_k \\
&~& \times \frac{(k-1)!}{\pi^{k-1}}
\frac{ (2\pi)^{k-2}}{2^{k-3}(k-3)!} \\
&=& 
 \frac{2 (k-1)(k-2)}{\pi}
\int_0^{2 \pi} \int_0^{\frac{\pi}{2}}\int_0^{\frac{\pi}{2}}
f(\theta_1, \theta_2, \theta_k ) 
\cos \theta_1 \sin^{2k-3} \theta_1 \,d \theta_1 
\cos \theta_2 \sin^{2k-5} \theta_2 \,d \theta_2 \,d \theta_k .
\end{eqnarray*}
Then we obtain (\ref{ha1}).
\end{pf}
\begin{lem}\label{hohe}
The following integral can be calculated as:
\begin{eqnarray}
\int_{0}^{1-a} x^m \log \left( \frac{1-x}{a} \right) \,d x
&=& \frac{1}{m+1}\left(
-\log a - \sum_{i=1}^{m+1}\frac{(1-a)^i}{i} \right) \label{ho}\\
\int_{0}^{\frac{a}{1+a}}
x^m \log \left ( \frac{x}{a(1-x)} \right) \,d x
&=& \frac{1}{m+1} \left( - \log (1+a) 
+ \sum_{i=1}^m \frac{1}{i} \left( \frac{a}{1+a} \right)^i \right). 
\label{he}
\end{eqnarray}
\end{lem}
\begin{pf}
The equation (\ref{ho}) is derived by the following:
\begin{eqnarray}
\int_0^{\alpha}  x^m \log (1-x) \,d x
= \frac{1}{m+1}\left(
(\alpha^{m+1} -1 ) \log ( 1- \alpha) 
- \sum_{i=1}^{m+1} \frac{\alpha^i}{i} \right). \label{ro}
\end{eqnarray}
Also, the equation (\ref{he}) is derived by (\ref{ro}) and the following:
\begin{eqnarray}
\int_0^{\alpha} x^m \log x \,d x
= \frac{1}{m+1}\left(
\alpha^{m+1} \left( \log \alpha - \frac{1}{m+1}\right) \right) .
\end{eqnarray}
\end{pf}
\begin{lem}\label{to}
We have the following equations:
\begin{eqnarray}
\sum_{i=0}^n {n \choose i} \frac{(-1)^i}{m+i}
=\int_0^1 x^{m-1}  (1-x)^n \,d x = 
{m+n \choose n}^{-1} \frac{1}{m}.
\end{eqnarray}
\end{lem}
It is easily derived.

\end{document}